# A study of the merging dwarf galaxy VCC322*

Lan-Yue Zhang,[1,2] Yinghe Zhao,[3,4] and Hong-Xin Zhang[5]

[1]*Yunnan Observatories, Chinese Academy of Sciences, Kunming 650011, China*
[2]*University of Chinese Academy of Sciences, Beijing 100049, China*
[3]*Yunnan Observatories, Chinese Academy of Sciences, Kunming 652016, China*
[4]*Key Laboratory of Radio Astronomy and Technology, Chinese Academy of Sciences, A20 Datun Road, Chaoyang District, Beijing, 100101, P. R. China*
[5]*CAS Key Laboratory for Research in Galaxies and Cosmology, Department of Astronomy, University of Science and Technology of China, Hefei, Anhui 230026, China*

## ABSTRACT

Galaxy interactions and mergers can enhance or reduce star formation, but a complete understanding of the involved processes is still lacking. The effect of dwarf galaxy mergers is even less clear than their massive counterpart. We present a study on a dwarf merger remnant in the Virgo cluster, VCC322, which might form a triple system with VCC334 and VCC319. We identify a prominent long and straight tail-like substructure that has a size comparable to its host galaxy VCC322. By comparing the color-color ($g - r$ vs. $r - H$) distribution with simple stellar population models, we infer that the metallicity and stellar age of this tail are $Z_\star \sim 0.02~Z_\odot$ and $t_\star \sim 10$ Gyr, respectively. In VCC319, we find a sign of isophotal twisting. This suggests that VCC319 may be subject to tidal interaction. An analysis of the SDSS optical spectra of VCC322 indicates mass- and light-weighted ages of about $10^{9.8}$ yr and $10^{7.5}$ yr, respectively, indicating an ongoing star formation activity. However, the star formation in VCC322 seems suppressed when compared to other star-forming dwarfs of comparable stellar masses. Our finding of shock excitation of optical emission lines indicates that interaction-induced shock may contribute to the heating of cold gas and suppression of star formation.

## 1. INTRODUCTION

According to the ΛCDM cosmological model, gravitational forces drive the ongoing growth and merging of structures. Visible galaxies grow continuously in mass through hierarchical mergers or accretion of smaller galaxies (Karachentsev et al. 2004; Hayashi & White 2008). Many observations and numerical simulations indicate that merges between massive galaxies can change galaxy morphologies (Genzel et al. 2001; Lotz et al. 2008; Bournaud et al. 2011), enhance star formation activities (Sanders & Mirabel 1996; Teyssier et al. 2010), and trigger starbursts and active galactic nuclei

Corresponding author: Yinghe Zhao; Hong-Xin Zhang
zhaoyinghe@ynao.ac.cn; hzhang18@ustc.edu.cn

* Released on XX, XX, XXXX

(Mihos et al. 1995; Patton et al. 2011, 2013; Torrey et al. 2012; Naab & Ostriker 2017; Cibinel et al. 2019).

When galaxies approach and pass through each other, tidal tails and bridges are produced by tidal forces and torques. These tidal structures can be used to trace for interacting events and merging events and retain some memory of the mass assembly of galaxies. Since the first imaging observations of collisional debris to exhibit the distinctive shapes of tidal tails and stellar streams (Arp 1966), some observations and numerical simulations have investigated the formation mechanism of tidal structures and their characteristic (Toomre & Toomre 1972; van Dokkum et al. 2014).

Dwarf galaxies ($M_* < 5 \times 10^9 M_\odot$) are the most numerous galaxies in the universe (Mateo 1998), and mergers between dwarf galaxies are expected to occur at all redshift (Klimentowski et al. 2010; Fitts et al. 2018). Observational challenges of dwarf mergers have resulted in a comparatively limited understanding of them, in contrast to their massive counterparts. Nevertheless, there is evidence that dwarf mergers play a significant role in triggering star formation in the nearby universe (Kravtsov 2013). The evolution of galaxy mergers may be different in low-mass end due to the shallow gravitational potential wells of dwarf galaxies. As a result, they are more susceptible to environmental influences and negative stellar feedback, which can significantly impact both the intensity and distribution of star formation (Geha et al. 2012; Kim et al. 2009).

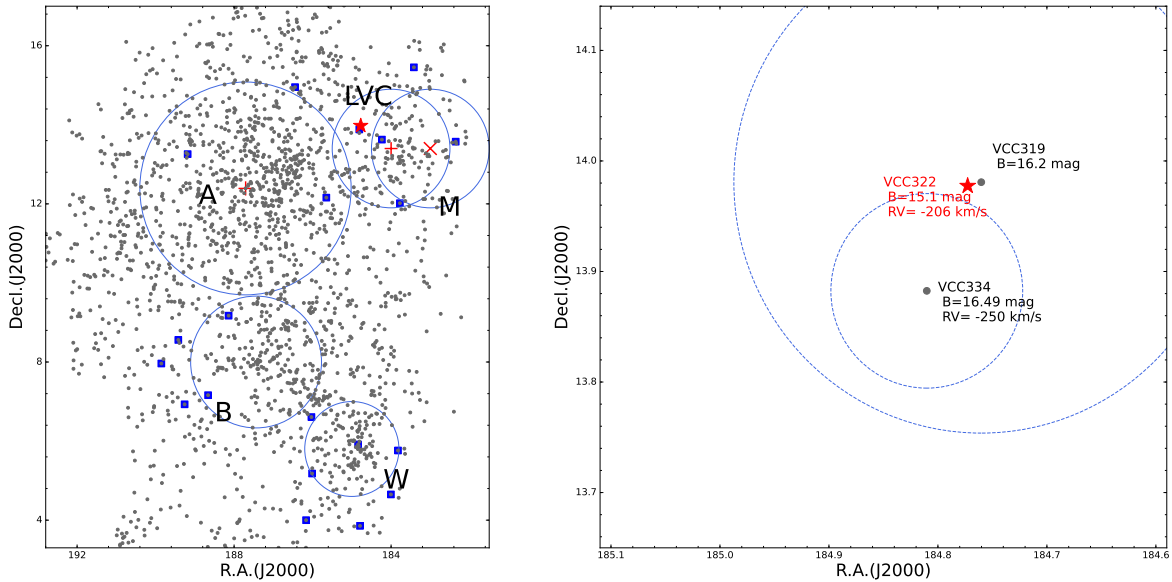

**Figure 1.** Location of VCC322 in the sky area of the Virgo cluster. The left panel presents the spatial distribution of VCC galaxies (gray symbols) (Binggeli et al. 1985). The red star symbol marks the location of VCC322. The two biggest blue solid circles mark half the virial radius of M87 (A) and of M49 (B) subclusters, respectively, and the three small black solid circles mark the boundaries of the M, LVC and W clouds, as defined in Boselli et al. (2014). The cross symbols and x symbol represent the central galaxy of the substructure. The blue square mark the location of a star formation dwarf galaxies sample from Grossi et al. (2016). The right panel is a zoom-in of the $0°.6 \times 0°.6$ sky area around VCC322. The blue dotted circles in the right panel mark half the virial radii of individual confirmed VCC galaxies, where the virial radius is approximated as the i-band half-light radius times 67 (Kravtsov 2013).

In recent years, there are some studies of dwarf-dwarf pairs (e.g. Stierwalt et al. 2015), merging dwarf systems (e.g. Paudel et al. 2018) and merging remnant (e.g. Zhang et al. 2020a). Stierwalt et al. (2015) found that, compared to isolated dwarf galaxies, the star formation rates (SFR) in paired dwarfs with a separation of $R_{\rm sep} < 50$ kpc are enhanced by a factor of $\sim 2$ on average. Moreover, this enhancement is found to decrease as $R_{\rm sep}$ increases. Nevertheless, Paudel et al. (2018) analyzed a sample of merging dwarfs selected based on their tidal structure features, and found that there is no significant SFR enhancement for their sources compared to a sample of local-volume, star-forming galaxies. They also found that the star formation properties of merging dwarf galaxies adjacent to massive galaxies are similar to those of isolated merging dwarf systems. Zhang et al.

(2020a,b) conducted the first comprehensive study of the assembly history of a blue compact dwarf merger remnant VCC848, and found that the formation rate of star clusters in the recent past have been significantly enhanced, but the current SFR is comparable to ordinary galaxies of similar masses. These studies suggests that the effect of dwarf mergers on star formation is complicated, and may depend various factors, such as the stage of merger.

In this paper we present a study on VCC322, a dwarf irregular galaxy located in the Virgo cluster, as shown in the left panel of Figure 1. It shows obvious tidal tails and stellar shells (see Figure 2), indicating its nature of a merger remnant. Furthermore, VCC322 might also be interacting with an early-type dwarf galaxy, VCC319, the occurrence of which is rare in galaxy clusters as compared to low-density environments (Paudel et al. 2018). Very few dwarf pairs involving early-type galaxies are in Stierwalt et al. (2015) and Paudel et al. (2018) samples. VCC322 (R.A.=184°.771465, Decl.=13°.980522) has a projected separation of about $\sim 0.5$ kpc from VCC319 (R.A.=184°.758261, Decl.=13°.982450). It has a radial velocity of -206.85 km $s^{-1}$. However, there is no reliable measurement of the radial velocity for VCC319 due to the lack of spectroscopic data. We will use imaging data to investigate the possibility that VCC322 and VCC319 being a physical pair in Section 3.

As the heliocentric velocity shows, VCC322 is a blue shifted dwarf galaxy, Karachentsev & Nasonova Kashibadze (2010) found that VCC322 and VCC334 form a pair with a small difference in radial velocity (right panel of Figure 1). VCC334 (R.A.=184°.809314, Decl.=13°.882401), which is ∼3.8 kpc away from VCC322, has a radial velocity of -250 km/s. Considering the fact that VCC322/334 lies within the half virial radius of VCC319 in projection (as illustrated in Figure 1), it is likely that VCC334/322/319 constitute a triple system. The neutral hydrogen mass of VCC322 is $M_{HI} = 1.86 \times 10^8 M_\odot$ (Haynes et al. 2011), and VCC322's HI deficiency parameter (the logarithmic difference between the measured HI mass and the HI mass of a reference sample of isolated galaxies with a given morphological type) is 0.31 (Grossi et al. 2015), which means that VCC322's atomic hydrogen gas fraction is about half that expected in the field environment (Haynes & Giovanelli 1984).

The paper is organized as follows. Section 2 describes the data reduction. We present our analysis of the phothometric and star-formation properties in Section 3 and summarize our results in the last section.

## 2. OBSERVATIONS AND DATA REDUCTION

The broadband optical images of VCC322 were retrieved from the Canadian Astronomy Data Center (CADC). The observations were performed by the Next Generation Virgo Cluster Survey (NGVS; Ferrarese et al. 2012) using the MegaCam instrument on the Canada France Hawaii Telescope (CFHT). The imaging reaches a $2\sigma$ (pixel-to-pixel noise) surface brightness limit of $\mu_{g,\rm lim} \simeq 26.2$ mag arcsec$^{-2}$ after combining multiple exposures[1], estimated with the photometric information given in the fits header. The $g$-band NGVS data have a pixel scale of $0''.186$ and a full width at half maximum (FWHM) of the point-spread function (PSF) of $\sim 0''.6$. In addition, $g$- and $r$-band images of VCC322 were also obtained by the Dark Energy Camera (DECam) on the Blanco 4m telescope, as part of The Dark Energy Camera Legacy Survey (Dey et al. 2019). These data reach a $2\sigma$ surface brightness limit of $\mu_g \sim 25.1$ mag arcsec$^{-2}$ and $\mu_r \sim 24.5$ mag arcsec$^{-2}$, and have a similar PSF size of FWHM $\sim 1''.1$. The near-infrared $H$-band image from the *Stellar content, MAss and Kinematics of Cluster Early-type Dwarf galaxies* project[2] (Janz et al. 2014) was also used in our analysis.

We perform structural measurements based on the exceptional deep CFHT $g$ band image. There are five exposures for the CFHT observations in the archive, with each exposure of ∼634 seconds. After downloading the calibrated data from the CADC website, we utilized IRAF for image reduction and combination, and performed background subtraction using the same approach as in Du et al. (2015).

VCC322 and VCC319 are close to each other in projection. To obtain clean measurement of the structures of the two galaxies separately, we first masked VCC322 and conducted surface photometry of VCC319 with Photutils package *photutils.isophote*. Then, we constructed a model image of VCC319 using the $g$-band surface brightness profile, and subtracted the model of VCC319 from the original image. The surface photometry of VCC322 was then carried out on the VCC319-model subtracted image. Note that the SExtractor package was used to generate mask of foreground/background sources when performing the surface photometry.

To assess color characteristics, we adopted the $g$- and $r$-band images from DESI survey, as we did not achieve an accurate flux calibration for the CFHT/MegaCam $r$-band images. We conducted aperture photometry on the target of interest, including VCC322, VCC319, and the tidal tails, using various aperture sizes. The aperture sizes of VCC322 (a=$34''.28$, e=1-b/a=0.17) and VCC319 (a=$58''.11$, e=1-b/a=0.27) were determined according to the $+1\sigma$ surface brightness contour of the DESI $g$-band image, whereas the apertures for the two tidal tails (Tail A and Tail B), as delineated in Figure 2, were determined using the $+2\sigma$ contour from the CFHT/MegaCam $g$-band image. Lastly, we correct for the Galactic extinction for the photometry by using the Schlegel et al. (1998) map. No attempt is made to correct for the internal extinction.

## 3. RESULTS AND DISCUSSION

### 3.1. *Surface Brightness Distribution and Photometry*

#### 3.1.1. *Tidal Features*

The left panel of Figure 2 shows the adaptively smoothed CFHT/MegaCam $g$-band image of VCC322/319, with a minimum S/N of 4. A straight tidal tail (hereafter Tail A; red polygon in the right panel of Figure 2) structure can be seen on the southeast side of the stellar main body, with a size comparable with that of the main galaxy. The tidal tail is oriented in the same direction as the major axis of VCC322. Towards the southeast of VCC322, a much smaller tail (Tail B), as shown by the blue polygon in the right panel of Figure 2, is almost perpendicular to the major axis. Based on the classification prescription described in Calderón-Castillo et al. (2019), VCC322/319 might be at stage IIIa (overlap, pre-merger) if they were a physical pair.

---

[1] This limit is different from that ($\mu_{g,\rm lim} = 29.0$ mag arcsec$^{-2}$) listed in Ferrarese et al. (2012), due to the fact that we have used the area of one pixel to measure $\mu_{g,\rm lim}$.

[2] https://dc.zah.uni-heidelberg.de/smakced/q/cat/info

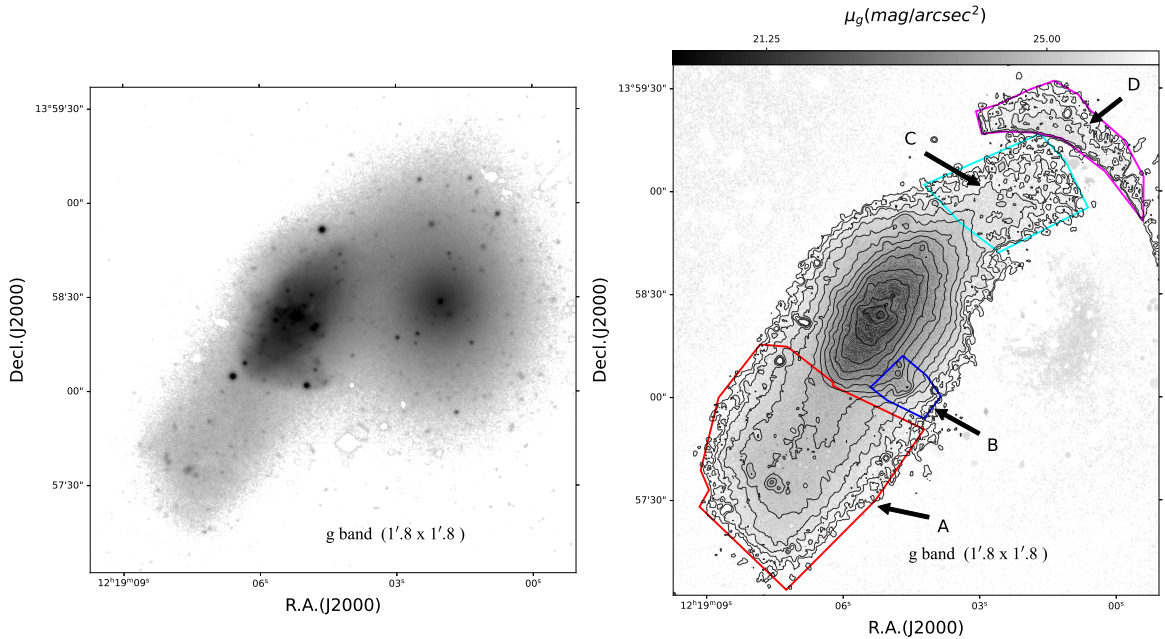

**Figure 2.** *Left*: $1'.8 \times 1'.8$ *g*-band image of VCC322/319 from CFHT/MegaCam. The original NGVS image has been adaptively smoothed to a minimum S/N of 4. *Right*: $1'.8 \times 1'.8$ *g*-band image of VCC322, obtained by subtracting the VCC319 model data and removing point sources. The contour is starting from background $+ 2\sigma$. The two tidal tails (named "Tail A" and "Tail B"), which are located in the south part of VCC322 and not contaminated by the light from VCC319, are shown by the red and blue polygons, respectively. A faint tidal tail (Tail C) and a faint tail-like structure (Tail D) can also be seen in the northwest of VCC322.

To have a clean view of VCC322, we subtracted the model image of VCC319 and masked point sources. The contours in the right panel of Figure 2 represent surface brightness levels ranging from $\sim 26$ mag arcsec$^{-2}$ to 21.25 mag arcsec$^{-2}$. From this figure we can also see a faint tidal tail (Tail C in Figure 2) in the northwest of VCC322, and a faint tail-like substructure (Tail D) to the west of VCC322 in the residual image.

### 3.1.2. *Isophotal Analysis*

The *g*-band stellar surface brightness profiles of VCC322 and VCC319 were derived by using the standard task *isophote* in Photutils, and are shown in Figure 3. Within the central $\sim 5''$ in radius, the ellipticity ($e$) and position angle (PA) of VCC322 vary significantly, whereas at larger radii they become stable, with $0.4 < e < 0.6$ and PA $\sim -40°$, indicating a highly disturbed stellar light distribution in the central region. The radial surface brightness profile within $R < 30''$ of VCC322 is fitted with a Sérsic function, and the best-fit parameters are $n = 0.98$ and $R_e = 12''$. The best-fit Sérsic profile is overplotted as a red curve in Figure 3. The obvious deviation from the best-fit profile at $R_{\rm maj} \gtrsim 25''$ is due to the presence of the above-mentioned long tidal tail (Tail A).

The ellipticity of VCC319 is less than 0.05 within the central $10''$ whereas it varies significantly as $R_{\rm maj}$ increases, as shown in the top right panel of Figure 3. From the middle right panel, we can see that VCC319 shows some variations in PA (i.e., isophotal twisting) at $R_{\rm maj} > 10''$, where the isophotes have significantly larger ellipticities than that at smaller radii. This indicates that VCC319 might be subject to tidal disturbance, possibly being induced by VCC322.

Using aperture photometry, we measured the *g*-, *r*- and *H*-band magnitudes, and thus the colors for the two galaxies and the two tidal tails, as listed in Table 1. The tidal tails have redder $g - r$ colors (0.57 and 0.44 mag respectively for Tail A and Tail B) compared to the host galaxy VCC322 (0.24 mag), but comparable $r - H$ colors. Whereas for VCC319, it has the reddest color.

To roughly estimate the stellar properties of these objects, we compare the above photometric results with the evolutionary synthesis models from Bruzual & Charlot (2003, hereafter BC03), as shown in Figure 4. Here we have chosen three single stellar populations (SSP) tracks with stellar metallicities of $Z_\star = 0.02$, 0.2 and 0.4 $Z_\odot$, computed using the Salpeter (1955) IMF, Padova-1994 models, and the STELIB library (Le Borgne et al. 2003). The choice of sub-solar abundances is because that low-mass galaxies are generally metal-poor according to the mass-metallicity ($M - Z$) relation (e.g., Tremonti et al. 2004; Zahid et al. 2017).

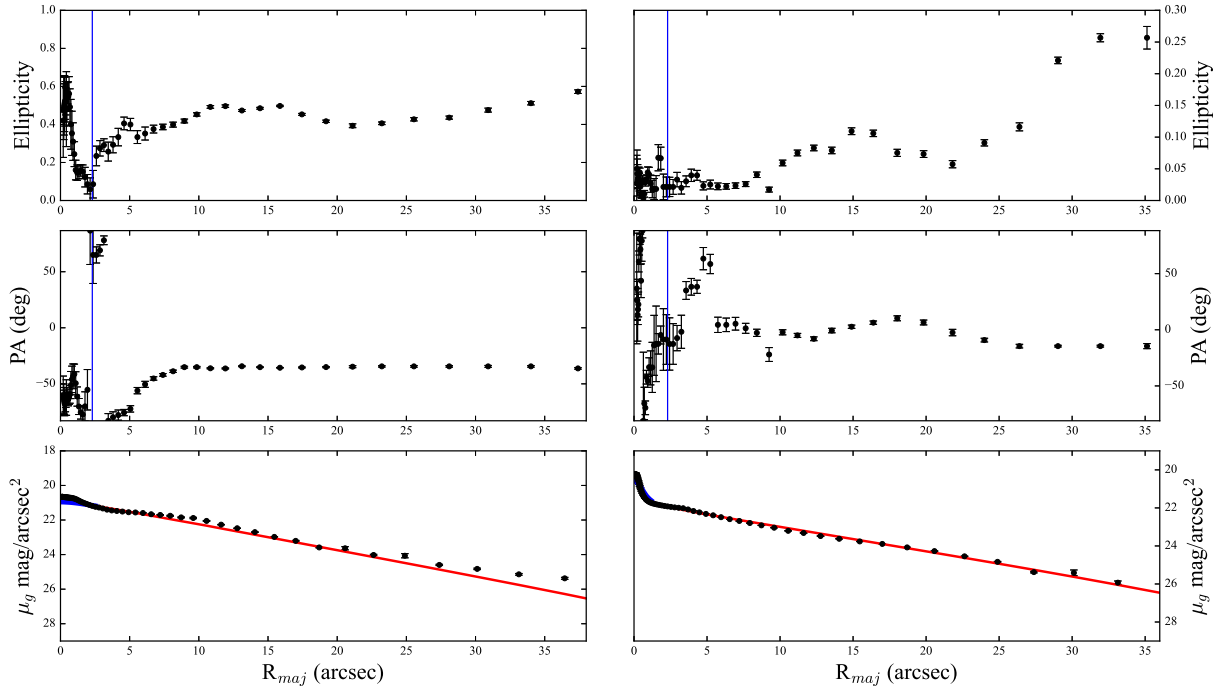

**Figure 3.** Results of ELLIPSE surface photometry of VCC322 (*left*) and VCC319 (*right*). From top to bottom are Ellipse, PA and surface brightness profile. The blue vertical lines represent $R_{maj}=2''.3$. The red line in the left panel represents the best fitted Sérsic profile (n=1, $R_h=13''$). The red line in the right panel represents the best fitted Gaussian + Sérsic profile (FWHM=$0.74''$, n=1.2, $R_h = 15''.1$).

From the figure we can see that Tail A is consistent with the SSP having $Z_\star = 0.02\ Z_\odot$ and a stellar age of $t_\star \sim 10$ Gyr, indicating there would be no star-forming activity occurred after its formation. For Tail B, its colors imply that it has a younger stellar age (and/or recent star formation) and a higher metallicity, which may be caused by the contamination of the host light since it is located within the outer part of VCC322. For VCC322, it agrees with the model tracks with metallicities of $0.2 - 0.4\ Z_\odot$, and has a younger luminosity-weighted stellar age. A more detailed analysis of the stellar populations will be presented in the following section. In contrast, VCC319 has a similar $t_\star$, consistent well with its early-type morphology, to Tail A but a much higher $Z_\star$ of $\sim 0.4\ Z_\odot$.

Such a low $Z_\star$ for Tail A is not unreasonable given the fact that its host galaxy, VCC322, has a very low (present) stellar mass of $1.3 \times 10^8\ M_\odot$ (see Table 1), which is estimated using the color-dependent mass-to-light ratio ($M_\star/L_r$) provided by Bell et al. (2003). Here we have converted the stellar mass to a Chabrier IMF by dividing a factor of 1.4. The derived mass of VCC322 is consistent with that estimated with the near-infrared (3.4 $\mu$m) luminosity (Grossi et al. 2015). According to the $M_\star-Z_\star$ relation in Zahid et al. (2017), a stellar mass of $1.3\times10^8 M_\odot$ implies a stellar metallicity of $\sim 0.13\ Z_\odot$, which should be treated as an upper limit for Tail A due to the metal enrichment caused by latter star formations in VCC322. For VCC319, the implied $Z_\star$ by the color-color diagram is also consistent with that ($\sim 0.26\ Z_\odot$) derived from Zahid 2017 $M_\star - Z_\star$ relation.

### 3.2. *Stellar Populations*

To obtain the stellar populations of VCC322, we follow Zhao et al. (2011) and Cai et al. (2020) using STARLIGHT (Cid Fernandes et al. 2005, 2007; Mateus et al. 2006; Asari et al. 2007) to fit the SDSS spectrum with a linear combination of $N_\star$ SSPs. The adopted base has $N_\star = 100$ with 25 ages from 1 Myr to 18 Gyr, and four metallicities of $Z_\star = 0.005, 0.02, 0.2,$ and $0.4\ Z_\odot$ for this metal-poor object, which were computed with the Salpeter IMF, Padova-1994 models, and the STELIB library using the evolutionary synthesis models from BC03. The fractional contribution of each SSP to the to-

**Table 1.** Photometric and Derived Properties

| Object | R.A. | Decl. | $m_r$ | $g-r$ | $r-H$ | n | $R_e$ | $\mu_{g,0}$ | $M_*$ | SFR | $M_{\rm H\,I}$ | z |
|---|---|---|---|---|---|---|---|---|---|---|---|---|
| | h:m:s | d:m:s | mag | mag | mag | | arcsec | mag arcsec$^{-2}$ | $M_\odot$ | $\log(M_\odot\,{\rm yr}^{-1})$ | $\log(M_\odot)$ | |
| VCC322 | 12:19:05.15 | 13:58:49.9 | 14.88 | 0.24 | 1.71 | 1 | 13 | 20.95 | $1.3\times10^8$ | -2.29 | 8.27 | -0.00069 |
| VCC319 | 12:19:01.9 | 13:58:56.8 | 14.68 | 0.77 | 2.29 | 1.2 | 15.1 | 20.54 | $5.2\times10^8$ | – | – | ... |
| Tail A | 12:19:06.9 | 13:57:47.9 | 16.51 | 0.57 | 1.68 | – | – | – | $5.9\times10^7$ | – | – | ... |
| Tail B | 12:19:04.7 | 13:58:02.7 | 18.19 | 0.44 | 1.84 | – | – | – | $9.4\times10^6$ | – | – | ... |

NOTE—$\mu_{g,0}$ was central surface brightness. The mass of H I was derived from the Arecibo Legacy Fast ALFA (ALFALFA) blind HI survey (Giovanelli et al. 2005), the $\alpha.40$ catalogue (Haynes et al. 2011). z was derived from NASA-Sloan Atlas (NSA) catalogue.

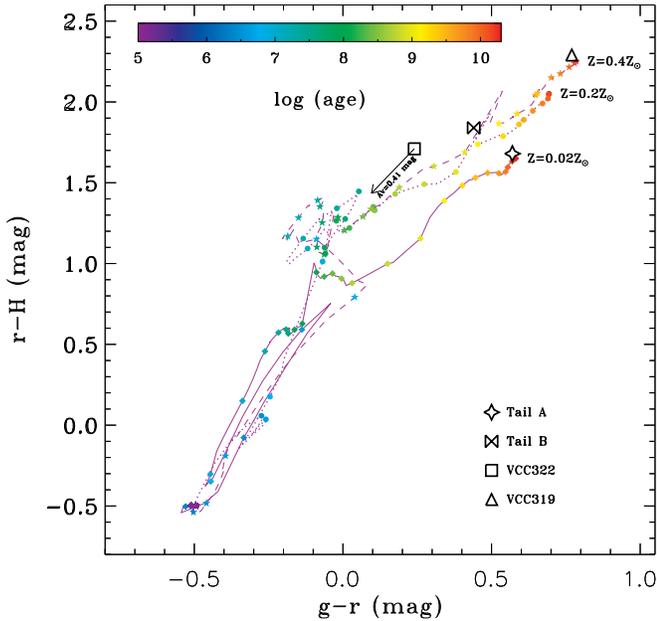

**Figure 4.** Observed $(g-r)$ plotted versus $(r-H)$ colors for the two dwarf galaxies VCC319 (open triangle) and VCC322 (open square), and the two tidal tails (Tail A: open four-angle star, and Tail B: Bowtie). The overplotted model tracks (solid symbols connected by lines) are single stellar populations of a Salpeter IMF with different ages (as indicated by the colorbar) and various stellar metallicities as annotated. The arrow demonstrates VCC322's position if it is corrected for the intrinsic extinction using $A_{V,\star} = 0.41$ mag.

tal synthetic flux is measured at the normalization wavelength $\lambda_0 = 4020$ Å.

Prior to the synthesis process, we correct the observed spectrum for redshift, and for Galactic extinction using the Cardelli et al. (1989) and O'Donnell (1994) Galactic extinction law with $R_V = 3.1$. Here we has adopted the $A_V$ value from Schlegel et al. (1998) through the NASA/IPAC Extragalactic Database (NED). To fit the observed spectrum having apparent emission lines, the intrinsic extinction $(A_{V,\star})$ is modelled with the foreground dust model and the Calzetti et al. (1994) extinction law with $R_V = 4.05$ Calzetti et al. (2000). During the fitting process, the SSPs are normalized at $\lambda_0$, and the observed spectrum is normalized to the median flux between 4010 and 4060 Å. The signal-to-noise ratio (S/N) of the observed spectrum is measured within the window of 4730 and 4780 Å, which is generally free of emission lines. The region of 20–30 Å is masked around obvious emission lines, and more weight is given to the strongest stellar absorption features (e.g., Ca II K $\lambda3934$, Ca II triplets) that are less affected by nearby emission lines. For more details about the synthesis process, please refer to Cid Fernandes et al. (2005) and Zhao et al. (2011).

The bottom panel of Figure 5 shows the false-color image of VCC322 from the DESI survey, overlaid with the aperture position of the SDSS spectrum. The upper left panel plots the observed (black line), synthetic (red line) and the pure emission-line (blue line) spectra, annotated with derived parameters from the best-fit result. As shown in the figure, VCC322 is experiencing a star formation activity with the light-weighted age ($\langle t_\star \rangle_L$) of $\sim$55 Myr, and the young population ($t_\star < 10^8$ yr) contributes most of the luminosity (upper right panel). Based on the stellar population results, the SFR averaged over the past 100 Myr is $\log({\rm SFR}_{100{\rm Myr,Chabrier}}) \sim -2.48\,M_\odot\,{\rm yr}^{-1}$. We can also derive the SFR using the non-dust corrected H$\alpha$ ($L_{{\rm H}\alpha}$) luminosity, exploiting equation (6) in Grossi et al. (2015), which is empirically re-calibrated from Lee et al. (2009) for low $L_{{\rm H}\alpha}$ sources based on FUV emission, and thus SFR$_{{\rm H}\alpha}$ traces the activity averaged over the past $\sim$100 Myr. From the pure-emission spectrum, we obtained $L_{{\rm H}\alpha} = 5.47 \times 10^{37}$ erg s$^{-1}$, and thus the derived $\log({\rm SFR}_{{\rm H}\alpha}) = -2.76 \pm 0.57\,M_\odot\,{\rm yr}^{-1}$, where the uncertainty is the 1$\sigma$ scatter between the H$\alpha$ and FUV SFRs, adopted from Table 2 of Lee et al. (2009). Therefore,

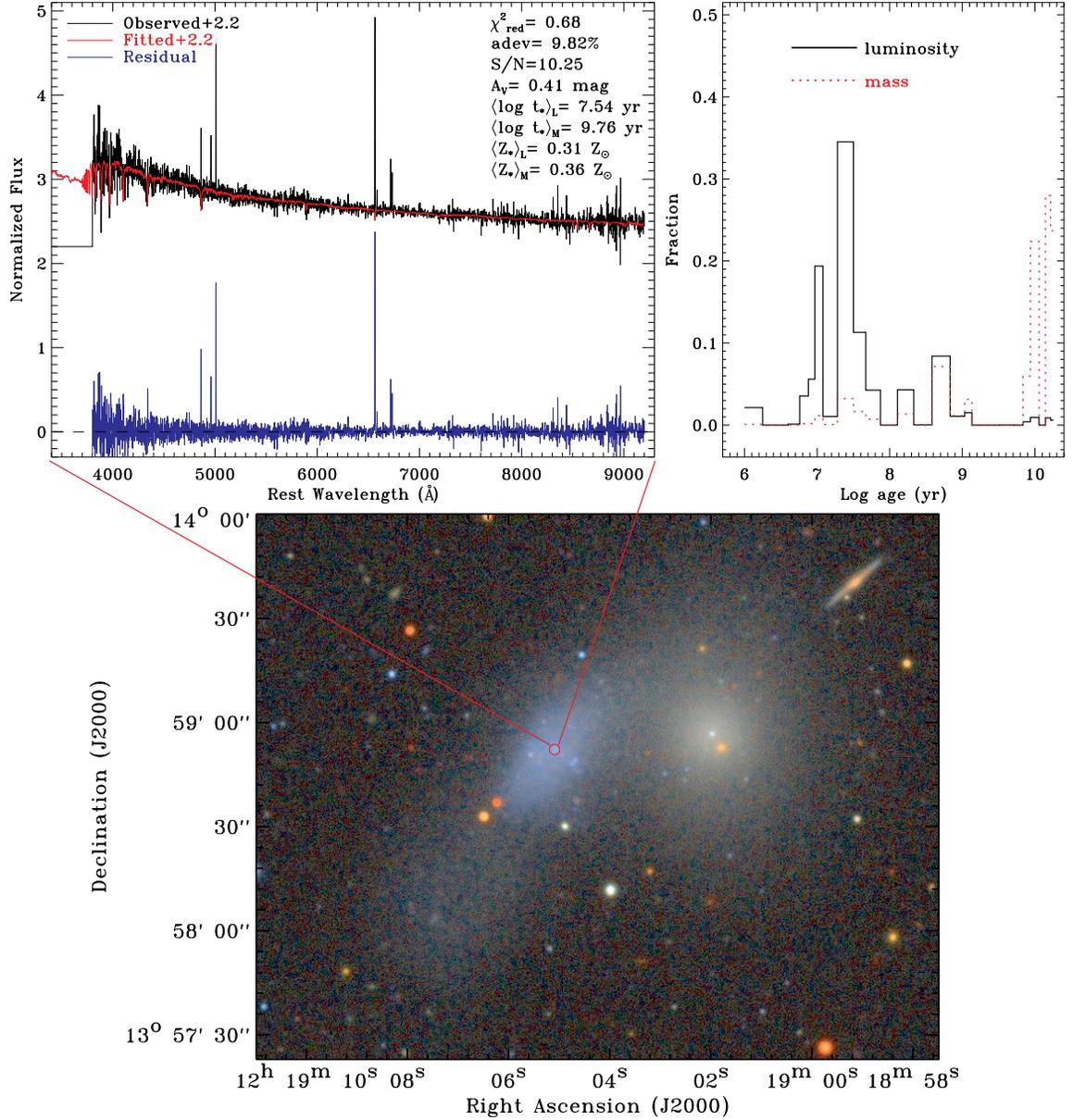

**Figure 5.** The best-fit spectrum for VCC322. The upper left panel plots the observed (black line), synthetic (red line) and the pure emission-line (blue line) spectra, annotated with derived parameters from the best-fit result. The upper right panel shows the distributions of luminosity (solid) and mass (dashed) fractions of different ages. The bottom panel is the false-color image of VCC322 from the DESI survey, overlaid with the aperture position of the SDSS spectrum.

these two SFRs are consistent with each other within uncertainties.

Besides the SFR, we can also derive the internal extinction using the Balmer line fluxes measured from the pure-emission spectrum. Assuming that the intrinsic Balmer line ratios are equal to Case B recombination, and adopting the Calzetti et al. (1994) reddening law, the derived nebular extinction in VCC322 is $A_{V,\mathrm{neb}} = 0.71$ mag, utilizing equation (5) in Zhao et al. (2011). Here we have adopted the intrinsic ratio of H$\alpha$/H$\beta$ to be 2.86 (Brocklehurst 1971) for an electron temperature of $10^4$ K and an electron density of 100 cm$^{-3}$. For the continuum, the extinction ($A_{V,\star}$) returned by STARLIGHT, as shown in Figure 5, is 0.41 mag, and thus $A_{V,\mathrm{neb}}/A_{V,\star} = 1.73$, approximately consistent with the finding in Calzetti et al. (1994) that nebular line emission is attenuated by roughly twice as much dust as the stellar continuum. These results indicate that the effect of the well-known age-extinction degeneracy (Gordon et al. 1997), which acts in the sense of confusing young, dusty systems with old, less dusty ones and vice versa, is not severe for the case of VCC322.

### 3.3. *Suppressed Star Formation?*

To explore the role of merger/interaction in the star-forming activity under dense environment, we further investigate the star formation properties of VCC322 by utilizing the SFR-$M_{\rm H\,I}$ (i.e. atomic hydrogen gas mass), SFR-$M_\star$ and SFR-to-$M_{\rm H\,I}$ ratio (i.e., the star formation efficiency; SFE), gas-to-stellar mass ratio ($M_{\rm H\,I}/M_\star$) relations, as shown in the left, middle and right panels of Figure 6, respectively. Here we adopt the SFR, $M_{\rm H\,I}$ and $M_\star$ from Grossi et al. (2015) for VCC322 and the star forming dwarfs (SFDs) in Virgo cluster. For comparison, we also plot the dwarf pairs ("I" class, i.e., ongoing interactions; 50 systems) and merger remnants ("T" class, i.e., showing tidal features; 13 systems) identified in Paudel et al. (2018). For the merging dwarf sample, the values of these physical parameters are given for the *total* system as in Paudel et al. (2018), and have been converted to the Chabrier IMF. Further, we have added 0.25 dex to the SFRs from Paudel et al. (2018) since they were derived with the FUV luminosity that has *not* been corrected for the internal dust extinction. This correction is estimated according to the median FUV attenuation ($A_{\rm FUV}$=0.6 mag; whereas mean $A_{\rm FUV}$=0.8 mag) for a sample of the local volume dwarf galaxies (Lee et al. 2009) having a similar range of $B$-band absolute magnitudes.

In each panel of Figure 6, the solid line is an unweighted least-squares linear fit, using a geometrical mean functional relationship (Isobe et al. 1990), to 19 Virgo SFDs having $N$ massive neighbours with $|N - N_{\rm VCC322}| \leq 5$, where $N_{\rm VCC322} = 20$. Here we have adopted the same criteria as Paudel et al. (2018) and utilize the NSA catalogue to determine the number of massive neighbours for both the Virgo SFDs and the dwarf merging samples. A galaxy is considered as a massive neighbour if it fulfills (1) $\log(M_\star/M_\odot) \geq 10$, (2) a sky-projected distance $|\Delta d| \leq 700$ kpc, and (3) a relative line-of-sight radial velocity $|\Delta v| \leq 700\,{\rm km\,s^{-1}}$. The two dashed lines are the corresponding $1\sigma$ deviation.

From figure 6 we can see that, albeit the large dispersion, SFR correlates with gas and stellar masses for Virgo SFDs and dwarf mergers, as found in more massive galaxies (Catinella et al. 2018). However, Virgo SFDs seem to have a different locus from merging dwarfs in the middle and right panels, i.e., at a given $M_\star$, merging dwarfs have a higher mean SFR. Virgo SFDs also show a lower SFE at a given gas-to-stellar mass ratio (right panel of figure 6). These differences might be due to the suppressed SF activity under dense environment and/or enhanced SF activity in merging/interacting systems. Regarding VCC322 (blue star in figure 6), it has a lower-than-average SFR and SFE among the Virgo SFDs. However, its companion VCC334 (red star in figure 6) has a much higher SFR and SFE, only slightly lower than the merging/interacting systems. Therefore, it seems that the SF activity in VCC322 is not enhanced but suppressed. In the following we will discuss the possible reasons caused this suppression.

Tidal stripping and ram pressure can strip atomic (e.g., Boselli & Gavazzi 2006) and molecular (e.g., Spilker et al. 2022) gas away from the host galaxy, which can suppress star formation and cause the galaxy to quench faster. Gas stripping may have occurred in the dwarf galaxy group consisting of VCC322/319/334, as suggested by the recent work (Bellazzini et al. 2018). The authors find that SECCO 1, an extremely dark, low-mass ($M_\star \sim 10^5\,M_\odot$) object located in the LVC substructure, has a similar oxygen abundance to the much more massive sources, VCC322/334, a feature of tidal dwarf galaxy (e.g., Duc et al. 2000). Therefore, it is very likely that SECCO 1 formed in a stripped gas cloud originating from the closest (about $\sim 250$ kpc away from VCC322) interacting triplet VCC322/319/334.

In a study of 10 pairs of gas-rich dwarf-pairs, Pearson et al. (2016) found that dwarf-dwarf interactions move gas to the outskirts of galaxies, with more than 50% of their total gas mass being beyond their 2MASS stellar extents. Pearson et al. (2018) subsequently used models to show that encounters between two dwarf galaxies can 'park' baryons at very large distances ($\sim 175$ kpc), even without the aid of environmental effects. The gas migration resulting from the merger, in conjunction with the ram pressure, provides assistance in stripping the gas (McPartland et al. 2016). If the H I gas of VCC322 is parked in the periphery by merge then stripped by ram pressure, the lower SFR of VCC322 is reasonable. However, without analysis of H I distribution, the contribution of tidal strength and ram pressure is unclear. With the interferometric data from VLA, we will discuss this effect in a future work (Zhang et al., in preparation).

Galaxy mergers can also shut down the star formation by heating up the gas supply via shocks, which consequently prevent it from gravitational collapse (e.g., Hopkins et al. 2008, 2009). Could the merging process of VCC322 trigger shocks that heating up the gas? To this end, we compared the observed [N II]$\lambda 6583$/H$\alpha$, [O II]$\lambda 6300$/H$\alpha$ and [S II]$\lambda\lambda 6716,6731$/H$\alpha$ ratios in VCC322 to the radiative shock models presented in Allen et al. (2008), as respectively shown in the left, middle and right panels of Figure 7. We have chosen the model grids that have an SMC abundance according to our stellar population synthetic results presented in section 3.2, and preshock density $n = 1$ cm$^{-3}$. Whereas for the [N II] $\lambda 6583$/H$\alpha$ plot, we have adopted the LMC abundance since the N/O ratio of SMC (log(N/O) = $-1.4$) is at the lower boundary for galaxies with similar metallicities (log(N/O) in the range of $-1.6$ and $-0.6$ for $12 + (\rm O/H) = 8 - 8.25$; Berg et al. 2019), and the modelled [N II] $\lambda 6583$/H$\alpha$ is too small to reproduce the observed value.

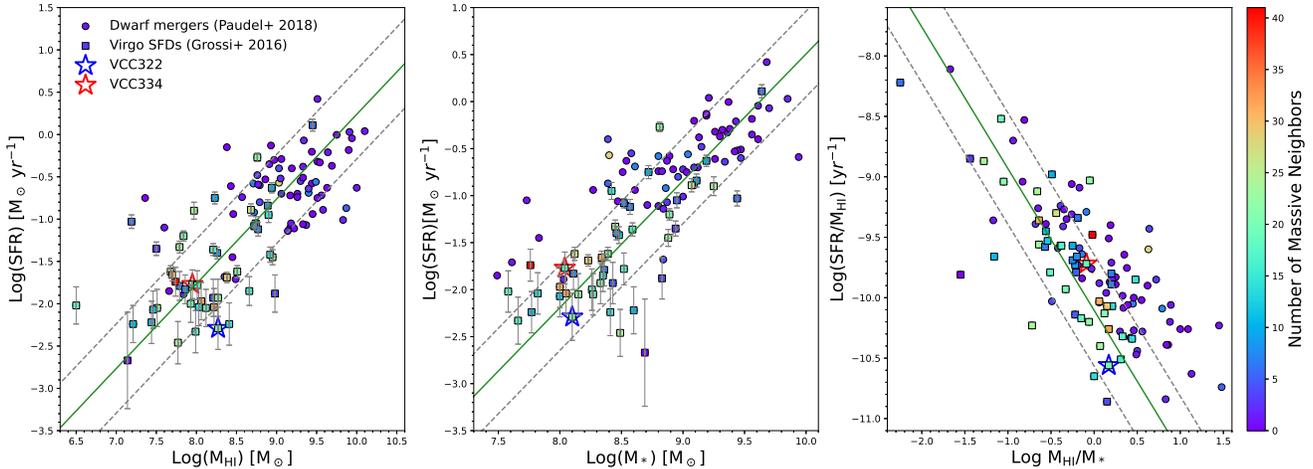

**Figure 6.** Relations between (1) SFR and gas mass ($M_{\rm H\,I}$; *left*), (2) SFR and stellar mass ($M_\star$; *middle*), and (3) SFE (SFR/$M_{\rm H\,I}$) and gas-to-stellar mass ratio ($M_{\rm H\,I}/M_\star$; *right*). We used the same symbols and colorbar in each panel. Blue and red stars represent VCC322 and VCC334, respectively; squares show the star-forming dwarf galaxies in Virgo cluster; solid circles plot the interacting dwarf pairs and merge remnants with tidal tails in P18; and colorbar displays the number of these galaxies' massive neighbors. In each panel, the (green) solid line shows the best-fit relation for galaxies having more than 20 massive neighbors, and the (gray) dashed lines represent $\pm 1\sigma$ deviation from the best-fit relation.

From figure 7 we can see that the observed line ratios fall in ambiguous regions defined by the Kauffmann et al. (2003) and Kewley et al. (2001) classification scheme. Based on the [N II] $\lambda 6583$/H$\alpha$ diagram, the optical line emissions are heated by the SF activity in VCC322, whereas from the [S II] $\lambda\lambda 6716,6731$/H$\alpha$ and [O II] $\lambda 6300$/H$\alpha$ diagrams an AGN is mainly responsible for the gas heating. However, this contradictory energy source might be reconciled with the help of the shock heating. A model with shock speed of $v_s = 200 - 250$ km s$^{-1}$ and magnetic field $B = 0.5 - 1\,\mu$G can well reproduce *each* observed line ratio, indicating that a merger/interaction-induced shock is highly likely to heat the gas and thus suppress the SF activity in VCC322.

In addition, Lisenfeld et al. (2019) find that Spiral+Elliptical pairs show no enhancement in SF and SFE. In their following work, Xu et al. (2021) suggest that Spiral+Elliptical pairs are more likely to experience high-speed and high-inclination interactions, which can generate ring-like density waves expanding through both stellar and gaseous disks, and thus pushing gas in the central region to the outer disk, resulting in a lower chance of high SFE nuclear starbursts. VCC322 and VCC319 form a Late+Early pair if they are physically bounded. However, we can not confirm whether these two sources are a high-speed interacting system due to the lack of reliable velocity measurement for VCC319. Regardless, the significantly low SFR and SFE in VCC322 may indicate that the effect found in Spiral+Elliptical pairs might also applicable to such kind of dwarf pairs.

## 4. SUMMARY

We conduct a comprehensive study on VCC322, a merger remnant which might form a triple system with VCC334 and VCC319, using multi-band photometric data and optical spectrum. We extracted its structural parameters, and explored its SF properties by performing SSP synthetics and comparing its SFE with other SFDs in Virgo cluster and merging dwarf sample. Our main results are:

1. The optical image of VCC322 reveals a prominent long straight tidal tail (Tail A), which has a size comparable to the host galaxy. The tidal tail has a $g-r$ color of 0.57 mag, with an inferred metallicity $Z_\star \sim 0.02\,Z_\odot$ and stellar age of $t_\star \sim 10$ Gyr.

2. The isophotal results show that VCC322 is a disk galaxy whereas VCC319 is an early-type object. Furthermore, VCC319 shows a sign of isophotal twisting, suggesting it is likely interacting with VCC322.

3. Based on the stellar population results, we obtain that the old population ($t_\star > 10^9$ yr) contributes most of the stellar mass of VCC322 with the mass-weighted age of $\sim 10^{9.8}$ year, whereas the young population ($t_\star < 10^8$ yr) contributes most of the luminosity with the light-weighted age of $\sim 10^{7.5}$ year.

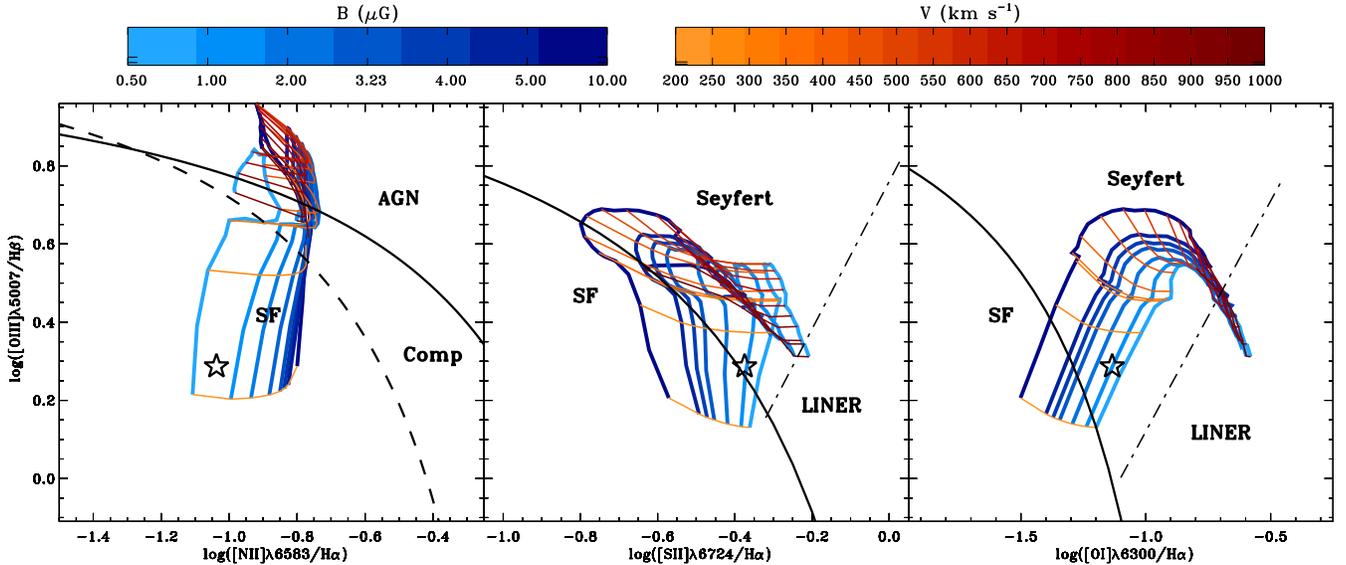

**Figure 7.** Comparison of the observed ratios to the fast shock model grids with an SMC abundance (LMC abundance for the [N II] $\lambda6583$/H$\alpha$ plot) and preshock density $n = 1$ cm$^{-3}$ (Allen et al. 2008) on the [N II] $\lambda6583$/H$\alpha$ (left), [S II] $\lambda\lambda6716,6731$/H$\alpha$ (middle) and [O I] $\lambda6300$/H$\alpha$ (right) vs. [O III] $\lambda5007$/H$\beta$ diagrams. The star shows the observed values in VCC322, and the thick (blue) and thin (orange) lines represent constant magnetic field and shock velocity, respectively. The continuous lines in the left panel separate AGNs, composite objects and star-forming galaxies (Kauffmann et al. 2003; Kewley et al. 2001). In the middle and right panels, the solid (black) line represents the Kewley et al. (2001) SF/AGN classification line, while the dot-dashed line shows the Seyfert-LINER division given by Kewley et al. (2006).

4. Comparing Virgo SFDs with dwarf merging systems, generally the latter has a higher SFE. However, VCC322 has a lower-than-average SFR and SFE among the Virgo SFDs, whereas its companion VCC334 has a much higher SFR and SFE, only slightly lower than the merging/interacting systems, indicating that the SF activity in VCC322 seems suppressed.

5. Comparing the optical emission line ratio with shock models, we suggest that a merger/interaction-induced shock ($v_s = 200 - 250$ km s$^{-1}$ and $B = 0.5 - 1\,\mu$G) is a probable cause of heating the gas and thus suppressing the SF activity in VCC322.

The authors thank the anonymous referee for valuable comments/suggestions. We acknowledge support from the China Manned Space Project (Nos. CMS-CSST-2021-A06 and CMS-CSST-2021-B02), the NSFC grant (Nos. 11421303, 11973039, 12122303 and 12173079), and the CAS Pioneer Hundred Talents Program, the Strategic Priority Research Program of Chinese Academy of Sciences (Grant No. XDB 41000000) and the Cyrus Chung Ying Tang Foundations. The STARLIGHT project is supported by the Brazilian agencies CNPq, CAPES, and FAPESP and by the France–Brazil CAPES/Cofecub program. This research made use of Photutils, an Astropy package for detection and photometry of astronomical sources (Bradley 2023). And LYZ thanks Si-Yue Yu for the enlightening photometry lesson at the CSST workshop. This research has made use of the NASA/IPAC Extragalactic Database (NED) which is operated by the California Institute of Technology, under contract with the National Aeronautics and Space Administration. All the authors acknowledge the work of the Sloan Digital Sky Survey (SDSS) team. Funding for SDSS-III has been provided by the Alfred P. Sloan Foundation, the Participating Institutions, the National Science Foundation, and the U.S. Department of Energy Office of Science. The SDSS-III web site is http://www.sdss3.org/. SDSS-III is managed by the Astrophysical Research Consortium for the Participating Institutions of the SDSS-III Collaboration including the University of Arizona, the Brazilian Participation Group, Brookhaven National Laboratory, Carnegie Mellon University, University of Florida, the French Participation Group, the German Participation Group, Harvard University, the Instituto